\documentclass[a4paper,11pt]{article}
\usepackage{jinstpub} 
\usepackage{lineno}

\title{\bf Upgrade of the Trigger and Data Acquisition System for Continuous Imaging and Multi-Camera Operation in CYGNO}

\author[1]{F.D.~Amaro}
\author[2,3]{R.~Antonietti}
\author[4,5]{E.~Baracchini}
\author[6]{L.~Benussi}
\author[6]{C.~Capoccia}
\author[6,7]{M.~Caponero}
\author[8]{L.G.M.~de~Carvalho}
\author[9,10]{G.~Cavoto}
\author[6]{I.A.~Costa}
\author[6]{A.~Croce}
\author[4,5]{M.~D'Astolfo}
\author[10]{G.~D'Imperio}
\author[6]{G.~Dho}
\author[10]{E.~Di~Marco}
\author[1]{J.M.F.~dos~Santos}
\author[4,5]{D.~Fiorina}
\author[10]{F.~Iacoangeli}
\author[4,5]{Z.~Islam}
\author[11]{E.~Kemp}
\author[4,5]{H.P.~Lima~Jr}
\author[6]{G.~Maccarrone}
\author[1]{R.D.P.~Mano}
\author[4,5]{D.J.G.~Marques}
\author[6]{G.~Mazzitelli}
\author[2,3]{P.~Meloni}
\author[9,10]{A.~Messina}
\author[9,10]{V.~Monno \footnote{Corresponding author.}}
\author[1]{C.M.B.~Monteiro}
\author[8]{R.A.~Nobrega}
\author[4,5]{G.M.~Oppedisano}
\author[8]{I.F.~Pains}
\author[6]{E.~Paoletti}
\author[2,3]{F.~Petrucci}
\author[4,5]{S.~Piacentini}
\author[6]{D.~Pierluigi}
\author[10]{D.~Pinci}
\author[10]{F.~Renga}
\author[6]{A.~Russo}
\author[6,12]{G.~Saviano}
\author[1]{P.A.O.C.~Silva}
\author[13]{N.J.~Spooner}
\author[6]{R.~Tesauro}
\author[6]{S.~Tomassini}
\author[9,10]{D.~Tozzi}

\affiliation[1]{LIBPhys, Department of Physics, University of Coimbra, 3004-516 Coimbra, Portugal}
\affiliation[2]{Dipartimento di Matematica e Fisica, Universit\`a Roma Tre, 00146 Roma, Italy}
\affiliation[3]{INFN Sezione di Roma Tre, 00146 Roma, Italy}
\affiliation[4]{Gran Sasso Science Institute, 67100 L'Aquila, Italy}
\affiliation[5]{INFN Laboratori Nazionali del Gran Sasso, 67100 Assergi, Italy}
\affiliation[6]{INFN Laboratori Nazionali di Frascati, 00044 Frascati, Italy}
\affiliation[7]{ENEA Centro Ricerche Frascati, 00044 Frascati, Italy}
\affiliation[8]{Universidade Federal de Juiz de Fora, Faculdade de Engenharia, 36036-900 Juiz de Fora, MG, Brasil}
\affiliation[9]{Dipartimento di Fisica, Sapienza Universit\`a di Roma, 00185 Roma, Italy}
\affiliation[10]{INFN Sezione di Roma, 00185 Roma, Italy}
\affiliation[11]{Universidade Estadual de Campinas (UNICAMP), Campinas 13083-859, SP, Brazil}
\affiliation[12]{Dipartimento di Ingegneria Chimica, Materiali e Ambiente, Sapienza Universit\`a di Roma, 00185 Roma, Italy}
\affiliation[13]{Department of Physics and Astronomy, University of Sheffield, Sheffield S3 7RH, UK}

\emailAdd{monno.1968992@studenti.uniroma1.it}

\abstract{
The CYGNO experiment employs an optical readout to image particle interactions in a gaseous
Time Projection Chamber (TPC), combining cameras and photomultiplier tubes (PMTs)
to achieve high spatial resolution and timing information. This approach enables detailed track reconstruction but poses significant challenges for data acquisition, particularly in view of the next experimental phase, CYGNO-04, which will operate multiple cameras simultaneously. In this paper, we present an upgrade of the CYGNO Trigger and Data Acquisition (T-DAQ) system, developed starting from the LIME configuration and validated on the MANGO prototype.
The upgrade introduces a continuous imaging acquisition mode, substantially reducing the
camera dead time, together with an extended trigger time-tagging scheme that provides a robust
global time reference for PMT signals.
A synchronous multi-camera DAQ architecture is also implemented and tested, enabling
coordinated operation of multiple optical sensors without a master camera.
The performance of the upgraded system is validated through dedicated tests, demonstrating
stable continuous acquisition, reliable time-tagging, and consistent synchronization across
multiple cameras.
These results establish a solid and scalable foundation for the CYGNO-04 DAQ and represent a
key step toward efficient data acquisition in future large-scale optical TPC detectors.}


\begin{document}
\maketitle
\flushbottom

\section{Introduction}
\label{sec:introduction}

Optical-readout Time Projection Chambers (TPCs) are increasingly employed in rare-event
searches~\cite{Mayet_2016zxu,Deaconu:2017vam}, where the reconstruction of short ionization tracks, typically extending over
$\mathcal{O}(10)$~mm, is crucial for background discrimination and directional sensitivity.
The CYGNO experiment adopts this approach~\cite{Amaro:2022gub,Vahsen:2020pzb} using a gaseous TPC operated near atmospheric
pressure with a He/CF$_4$ gas mixture, in which charge amplification in a triple-GEM \cite{Sauli1997, bib:Fraga:2003uu} stack
produces scintillation light.
This light is imaged by scientific cameras equipped with Active Pixel Sensor (APS) based on CMOS technology producing megapixel-scale images, while photomultiplier tubes (PMTs) record the prompt optical signal, providing complementary timing information.
This readout strategy, while highly effective, poses specific challenges for the Trigger and
Data Acquisition (T-DAQ) system.
The cameras operate with exposure times in the range of tens to hundreds of
milliseconds, producing megapixel-scale images of the full active area, whereas PMTs work in the nanosecond scale.
Efficiently integrating these heterogeneous data streams, while maintaining precise timing
and minimizing dead time, is a central requirement for ensuring high data quality and reliable event reconstruction.

The CYGNO collaboration has progressively developed and tested this technology through a
series of prototypes, culminating in the underground operation of the LIME detector~\cite{LIME}.
The forthcoming CYGNO-04 demonstrator will represent a significant step forward, employing
multiple optical readout units for the first time and requiring sustained, synchronized
acquisition from several cameras operating simultaneously.
This evolution places stringent requirements on the T-DAQ system in terms of dead-time
reduction, timing robustness, and scalability.

In this context, the work presented in this paper focuses on an upgrade of the CYGNO T-DAQ
system developed starting from the LIME configuration and validated on the MANGO prototype,
a small-scale optical TPC used for CYGNO R\&D studies~\cite{Benussi2020_HeCF4firstevidenceOFluminescence,Amaro2024_Enhancingthelightyield}.
The upgrade addresses the main operational requirements for CYGNO-04 by introducing
continuous camera acquisition to minimize dead time and by enabling synchronous operation
of multiple cameras within a unified DAQ architecture.

In addition, the trigger time-tagging strategy is revised to provide a global time reference
for PMT signals over extended data-taking periods, ensuring unambiguous temporal association
between camera and PMT information.
The performance of the upgraded T-DAQ system is validated through dedicated tests, showing
stable continuous acquisition, reliable timing, and consistent synchronization across multiple
cameras.

This paper describes the design, implementation, and validation of these upgrades and
discusses their implications for the CYGNO-04 detector, establishing a solid foundation for the
next phase of the CYGNO experiment.

\section{The CYGNO LIME-Based T-DAQ System}
\label{sec:lime_daq}

The CYGNO experiment employs, in the LIME prototype \cite{LIME}, a T-DAQ system originally developed to integrate the two readout channels of the optical TPC: a camera and four PMTs.
Each recorded event therefore consists of a camera image and a corresponding set of digitized
PMT waveforms. A defining feature of the system is the coexistence of two intrinsically different
time scales: PMTs work in the nanosecond scale, while the camera exposure time,
in low-noise configurations, is typically in the range of tens to hundreds of milliseconds.
This section reviews the architecture and operating principles of the LIME T-DAQ system and
identifies the limitations that motivated the upgrade described in the following sections.

\subsection{LIME DAQ architecture and trigger logic}
\label{subsec:lime_architecture}

\begin{figure}[h!]
    \centering
    \includegraphics[width=0.85\linewidth]{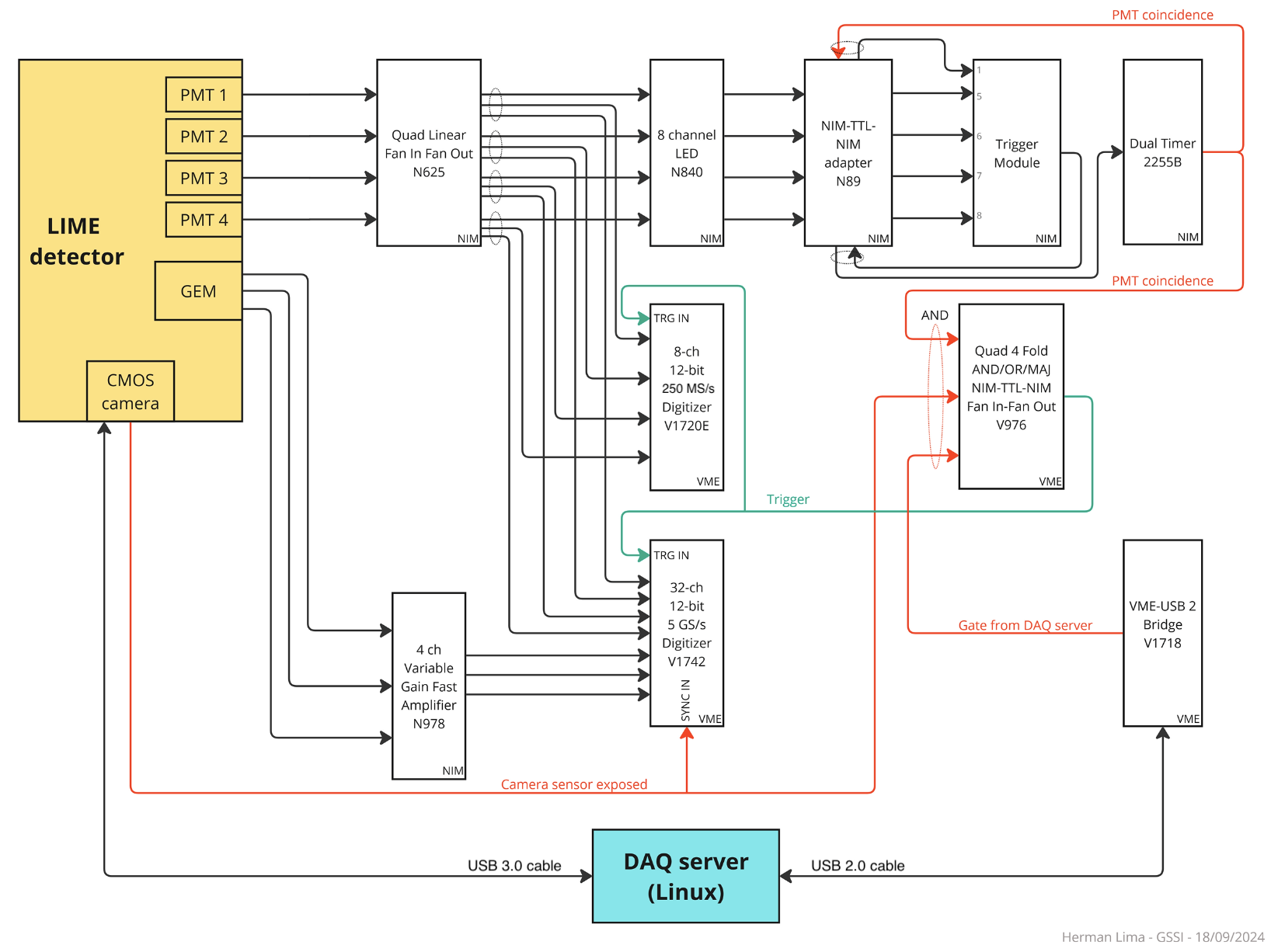}
    \caption{Complete scheme of the LIME DAQ employing NIM and VME modules. The delivered trigger is shown in green, while the coincidence signals producing it are shown in red.}
    \label{fig: DAQlimeScheme}
\end{figure}

The LIME T-DAQ architecture is built around a single camera operated in externally
triggered mode and four PMTs providing fast optical signals from the TPC.
A schematic overview of the system is shown in Fig.~\ref{fig: DAQlimeScheme}.
The acquisition flow is governed by a tight coupling between camera exposure and PMT-based
trigger generation, implemented through a combination of NIM and VME modules and coordinated
by the \texttt{MIDAS} data-acquisition framework \cite{MIDAS}.

\begin{figure}
    \centering
    \includegraphics[width=0.95\linewidth]{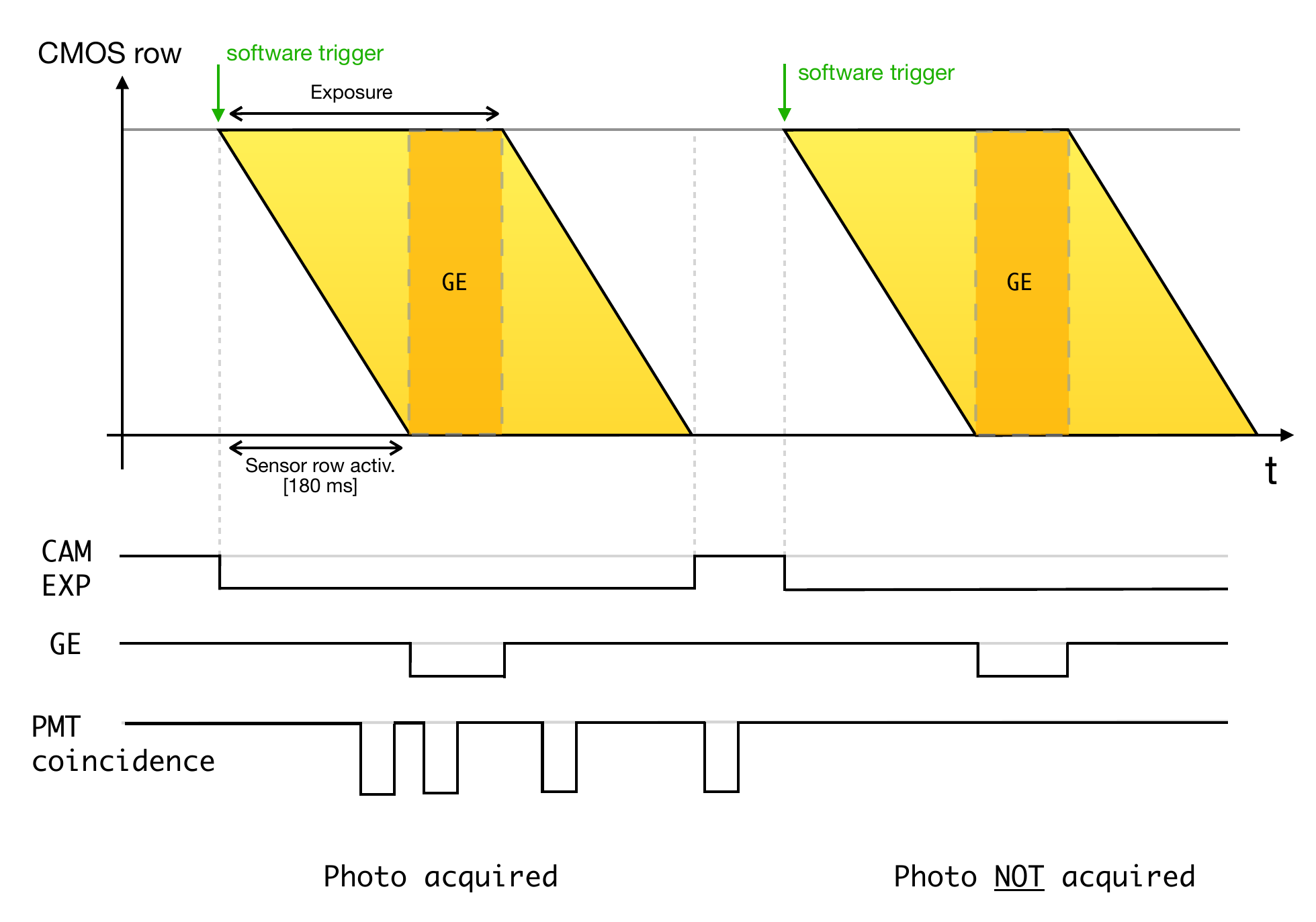}
    \caption{Timing diagram of the LIME camera acquisition. The software trigger starts the exposure of the CMOS rows, during which the global exposure (GE) occurs when all rows are simultaneously active. The corresponding \texttt{CAMEXP} and \texttt{GE} NIM logic signals are shown in the lower traces. The diagram also reports the PMT coincidence signal. In the second frame shown, no PMT coincidence occurs within the camera-sensitive interval and the camera image is therefore not acquired. }
    \label{fig:FrameBasedTiming}
\end{figure}

The DAQ sequence is initiated by a software trigger issued by the DAQ server to a Hamamatsu ORCA-Fusion CMOS-based camera~\cite{HamamatsuOrcaFusionManual}.
Upon receiving this trigger, the camera starts exposing its sensor rows sequentially.
Since the activation of all rows is not instantaneous, the interval during which the full sensor
area is simultaneously sensitive, the global exposure (GE), is shorter than the nominal exposure
time \(T_\mathrm{exp}\) Fig.~\ref{fig:FrameBasedTiming}, defined as the programmed exposure duration applied to each individual sensor row. For the LIME configuration considered here, the activation time of all sensor rows is
\(T_\mathrm{act} \simeq 184.4~\mathrm{ms}\).
The camera provides a programmable digital output signal, denoted as \texttt{CAMEXP}, which is
asserted whenever at least one sensor row is active. This signal is exported as a NIM logic level
and defines the temporal window during which physical interactions can be accepted.
The timing sequence of the camera operated in externally triggered frame-based acquisition mode,
together with the associated \texttt{CAMEXP} signal, is illustrated in Fig.~\ref{fig:FrameBasedTiming}.

The \texttt{CAMEXP} signal is combined in logical \texttt{AND} with the PMT coincidence signal,
thereby gating the trigger acceptance to the camera-sensitive interval.
In addition, \texttt{CAMEXP} is fed to the PMT digitizer as a synchronization input, ensuring a
well-defined temporal reference between the camera exposure and the trigger time tag associated
with the PMT waveforms.
This scheme enforces a strict correspondence between camera exposure and PMT trigger
generation.

Each of the four PMT analog signals is split using a linear fan-in fan-out module.
Two copies of each signal are routed to the digitizers for waveform acquisition, while an
additional copy is processed by a leading-edge discriminator.
PMT waveforms are acquired in parallel by a 32-channel 12-bit 5~GS/s digitizer (\texttt{CAEN V1742}\cite{CAENV1742}), operated at a sampling frequency of 750~MS/s, providing high temporal resolution over a short acquisition window ($\sim$1.4~$\mu$s), and by an 8-channel 12-bit digitizer (\texttt{CAEN V1720E}\cite{CAENV1720}), operated at 250~MS/s, which offers a longer acquisition window ($\sim$16~$\mu$s) sufficient to cover the full drift time of the TPC.
The discriminator outputs are fed into a custom trigger module, which generates a PMT
coincidence when at least two channels exceed the programmed threshold within a predefined
time window.
The final trigger signal delivered to the PMT digitizers is obtained by combining, in logical
\texttt{AND}, the PMT coincidence, the \texttt{CAMEXP} gate, and an open-gate signal ensuring
that the digitizers are not busy.
Upon trigger acceptance, PMT waveforms are acquired, while the corresponding camera image
is transferred to the DAQ server via a direct USB~3.0 connection.

The data acquisition and event building are handled within the \texttt{MIDAS} framework.
Hardware devices are interfaced through dedicated front-end processes, while configuration
parameters are stored in the Online Database (ODB).
Data are written to disk in the MIDAS format as a sequence of events, each organized into
structured data banks containing camera images, PMT waveforms, and associated metadata.
This structure enables a straightforward offline reconstruction and subsequent processing.

\subsection{Limitations of the frame-based acquisition scheme}
\label{subsec:lime_limitations}

A critical limitation of the LIME T-DAQ system arises from the frame-based camera acquisition,
which introduces a non-negligible dead time.
The first contribution to the dead time is intrinsic to the camera duty cycle.
The total duration of a frame is given by the sum of the sensor rows activation time and the
exposure time,
\begin{equation}
T_\mathrm{frame} = T_\mathrm{act} + T_\mathrm{exp}.
\end{equation}
For a typical LIME configuration, the exposure time is set to \(T_\mathrm{exp} = 300~\mathrm{ms}\), representing a compromise between noise integration
and data throughput: longer exposures reduce the relative dead time but increase the
contribution of time-independent noise, while shorter exposures limit noise at the cost of
higher frame rates and increased data storage requirements.
Together with a sensor row activation time of
\(T_\mathrm{act} \simeq 184.4~\mathrm{ms}\), this yields
\begin{equation}
T_\mathrm{frame} \simeq 484~\mathrm{ms}.
\end{equation}
Since only the exposure interval corresponds to an active acquisition window, the resulting
dead-time fraction is
\begin{equation}
f_\mathrm{DT}
= 1 - \frac{T_\mathrm{exp}}{T_\mathrm{frame}}
\simeq 38\%.
\end{equation}
This inefficiency is intrinsic to the externally triggered frame-based operation and cannot be
mitigated without modifying the acquisition paradigm.

A second contribution to the dead time originates from the requirement that the digitizer
readout must be completed before a new software trigger can be issued to the camera.
This introduces an additional inactive gap between consecutive frames, further reducing the
overall live time.
Together, these effects significantly limit the acquisition efficiency.
This is particularly relevant for low-energy and localized events, which typically extend over only one or a few sensor rows.
Such events can be entirely lost if they occur during a row-level inactive interval.

In addition, the frame-based acquisition scheme can lead to partial recording of extended tracks.
Ionization electrons produced may reach the optical readout during periods in which a fraction of the sensor rows are not actively acquiring, resulting in tracks that are truncated.
This effect complicates event reconstruction and can bias the measurement of track topology.
These limitations motivated the adoption of a continuous imaging acquisition mode in the
upgraded T-DAQ system, which removes the inactive interval within a frame and decouples the
camera exposure logic from the PMT readout.
The design and validation of this upgrade are presented in the next section.

\section{Continuous Imaging T-DAQ Upgrade}
\label{sec:continuous_imaging}

\subsection{Design Requirements and Concept}
\label{subsec:ci_concept}

The primary design requirement of the upgraded T-DAQ was the elimination of the
inactive interval within each camera frame, while preserving the existing hardware configuration
and maintaining full backward compatibility with the LIME setup.
This led to the adoption of a continuous imaging paradigm, in which the camera exposure is
decoupled from the trigger logic and proceeds uninterrupted for the full duration of a run.

\subsection{Continuous Imaging Implementation}
\label{subsec:ci_implementation}

\begin{figure}[h!]
  \centering
  \includegraphics[width=0.95\linewidth]{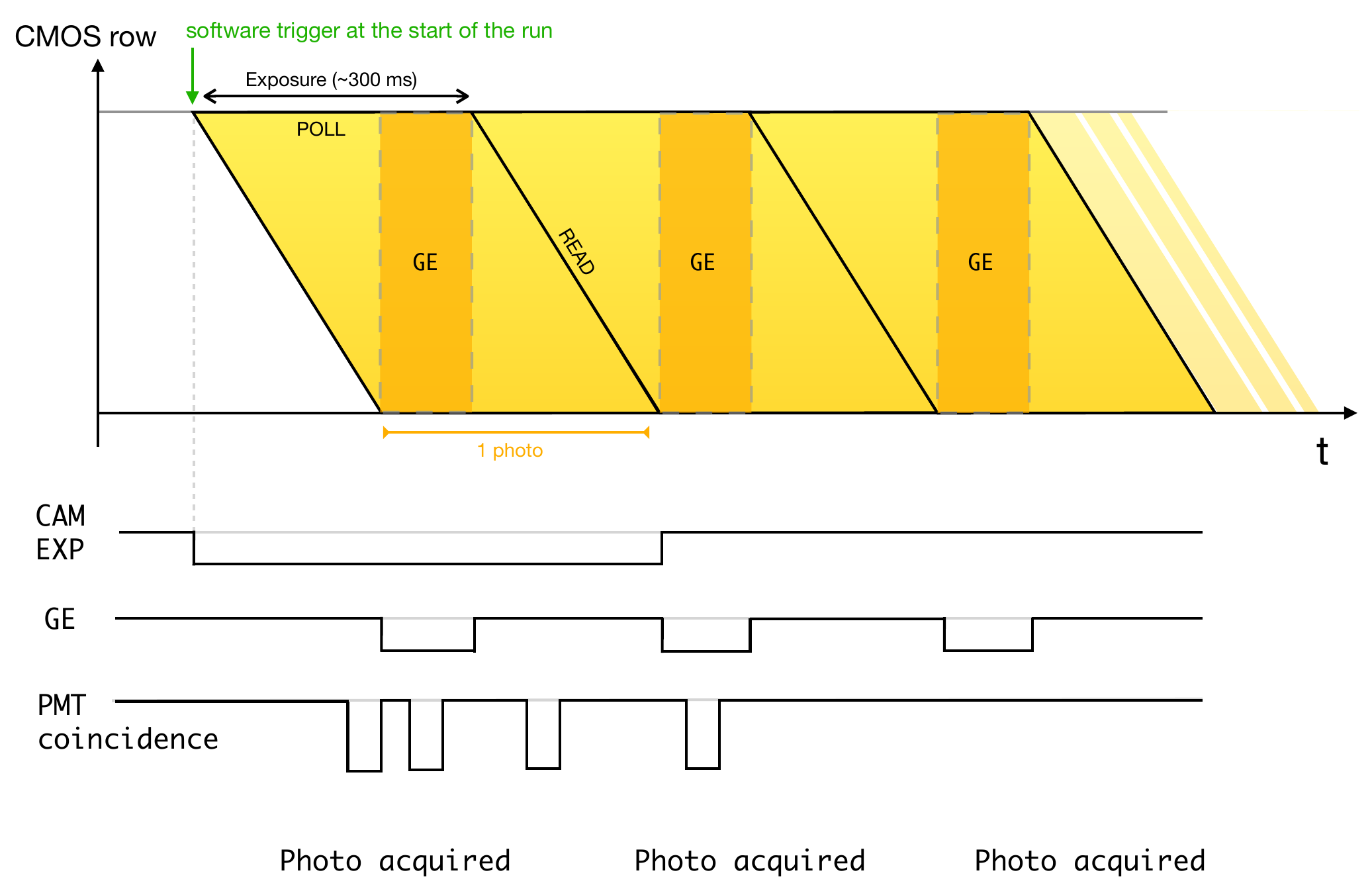}
  \caption{Timing diagram of the camera operated in continuous acquisition mode.
  The camera is started by a single software trigger at the beginning of the run and subsequently
  acquires frames continuously.
  The exposure proceeds without interruptions on a frame-by-frame basis, resulting in a
  negligible dead-time fraction.}
  \label{fig:TimeDiagramContinuous}
\end{figure}

Continuous imaging is implemented by operating the camera in \texttt{START TRIGGER}
mode.
At the beginning of each run, a single software trigger is issued to the camera, which then
acquires frames continuously until the run is stopped.
Unlike the frame-based operation, no further triggers are required to control the exposure
sequence.
In this configuration, the camera exposure signal (\texttt{CAMEXP}) is generated only for the
first acquired frame and is therefore no longer informative on a per-frame basis.
Subsequent frames are acquired back-to-back, with only a minimal interruption associated with
the readout and reset of individual sensor rows.
The timing diagram of the continuous acquisition mode and the associated logic signals are illustrated
in Fig.~\ref{fig:TimeDiagramContinuous}.

An important consequence of this operating mode is the dramatic reduction of the dead-time
fraction.
For a representative configuration with an exposure time of \(T_\mathrm{exp}=300~\mathrm{ms}\)
and a row readout time of \(t_\mathrm{row}=86.4~\mu\mathrm{s}\), the inactive fraction of each frame
is limited to the readout of a single row.
The corresponding dead-time fraction is therefore
\begin{equation}
f_\mathrm{DT}
= \frac{t_\mathrm{row}}{T_\mathrm{exp}+t_\mathrm{row}}
\simeq 0.03\%,
\end{equation}
which is negligible when compared to the \(\sim38\%\) estimated for the frame-based acquisition.
Moreover, since the dead time affects only one row at a time, the probability of losing a full
extended track is strongly suppressed, as the remaining sensor rows remain active or have
already been read out.

\subsection{Extended Group Trigger Time Tag}
\label{subsec:egttt}

Operating the camera in continuous mode requires a revision of the PMT time-tagging strategy.
In the original LIME configuration, the Trigger Time Tag (TTT) provided by the CAEN V1742
digitizer was reset for each camera exposure and referenced to the \texttt{CAMEXP} signal.
In continuous imaging, however, a single physical event may occur during the transition between
two consecutive frames, and the readout of the camera and PMTs is no longer synchronized on a
per-frame basis.

To avoid ambiguities in the temporal association between PMT signals and camera images, PMT
events are therefore time-stamped with respect to a common reference corresponding to the
beginning of the run.
In this configuration, the standard 30-bit TTT provided by the digitizer, with a maximum range
of approximately 9.1~s, becomes insufficient for typical run durations.

To overcome this limitation, the Extended Group Trigger Time Tag (EGTTT), available in the CAEN V1742 digitizer firmware \cite{CAENV1742}, was enabled by upgrading to a firmware version supporting this feature.
The EGTTT consists of a 60-bit time tag formed by combining two 30-bit registers, providing a
significantly extended dynamic range while preserving compatibility with the existing trigger
logic.
The least significant 30 bits correspond to the original TTT, while the most significant bits
extend the timestamp beyond the standard rollover limit.
The combined EGTTT value is reconstructed offline using standard bit-wise operations, and the
data format has been updated accordingly to ensure transparent access within the CYGNO
software framework.
No modifications to the digitizer firmware were performed; the EGTTT functionality was used as provided by the manufacturer and integrated into the DAQ data format and reconstruction workflow.

\subsection{Performance Validation}
\label{subsec:ci_validation}

The continuous imaging T-DAQ upgrade was validated through a series of dedicated tests aimed
at verifying the integrity and timing consistency of the acquired data.
The DAQ was confirmed to correctly handle the continuous camera acquisition, with no skipped or duplicated frames observed in the recorded data during dedicated runs.
In a representative test, a sequence of 1039 consecutive frames was acquired over a duration
of approximately 6~min and 46~s, with no skipped or duplicated images observed.
The camera-provided timestamp exhibits a uniform progression consistent with the configured
exposure time.
The distribution of timestamp differences, evaluated with a binning of $1~\mu$s corresponding
to the hardware resolution, shows two discrete values with a peak-to-peak spread of
$1~\mu$s.
This behavior indicates that the observed jitter is limited by the intrinsic timestamp
granularity and confirms the temporal stability of the continuous acquisition.
In addition, the camera frame index was verified to increase monotonically throughout the run,
demonstrating the absence of frame losses at the DAQ level.\\
On the PMT side, the correct operation of the trigger logic and the extended time-tagging was
validated using periodic test signals.
The EGTTT was observed to extend the usable timestamp range well beyond the standard
30-bit limit, enabling unambiguous time stamping of PMT events over the full duration of a run, which typically spans from a few minutes up to \(\mathcal{O}(10)\) minutes.
Together, these tests confirm the reliable operation of the DAQ in continuous imaging mode and its capability to correctly acquire and time-stamp data, providing a robust foundation for subsequent developments, including multi-camera synchronous acquisition.

\section{Multi-Camera Synchronous DAQ Architecture}
\label{sec:multicamera_daq}

The upgrade to a multi-camera synchronous data acquisition represents a crucial step toward
the realization of the CYGNO-04 detector, which will employ multiple optical readout units for
the first time in the CYGNO experiment.
The solution developed and tested on the MANGO prototype, employing Hamamatsu QUEST~2 cameras~\cite{HamamatsuQuest2Manual}, was designed to extend the continuous imaging paradigm to multiple cameras, ensuring synchronous operation while preserving modularity and scalability.
This section describes the synchronization strategy, the hardware logic adopted for timing
distribution, and the validation of the multi-camera acquisition.

\subsection{Synchronization Strategy}
\label{subsec:multicamera_strategy}

The primary requirement for multi-camera operation is the ability to acquire images from
different cameras with a well-defined and reproducible temporal relationship.
To achieve this, the cameras are operated in \emph{synchronous readout trigger mode} \cite{HamamatsuQuest2Manual}, in which
frame exposure and readout are driven by a common external timing signal.
In this mode, all cameras start exposing their first frame
on the edge of the same trigger pulse, and subsequent frames are triggered after a predefined number
of pulses, ensuring synchronous exposure cycles.

The system architecture is intentionally designed without a master camera, i.e. without a single camera providing trigger signals to the others.
In principle, any camera can be selected to provide the signals required for synchronization,
allowing flexible operation and improved fault tolerance.
This feature is particularly relevant for CYGNO-04, where subsets of cameras may be operated
independently during commissioning, calibration runs, or in the presence of partial detector
failures.
To support this flexibility, a configurable camera selection mask has been implemented at the
DAQ software level.

\subsection{Hardware Logic and Timing Distribution}
\label{subsec:multicamera_hardware}

\begin{figure}[]
  \centering
  \includegraphics[width=0.95\linewidth]{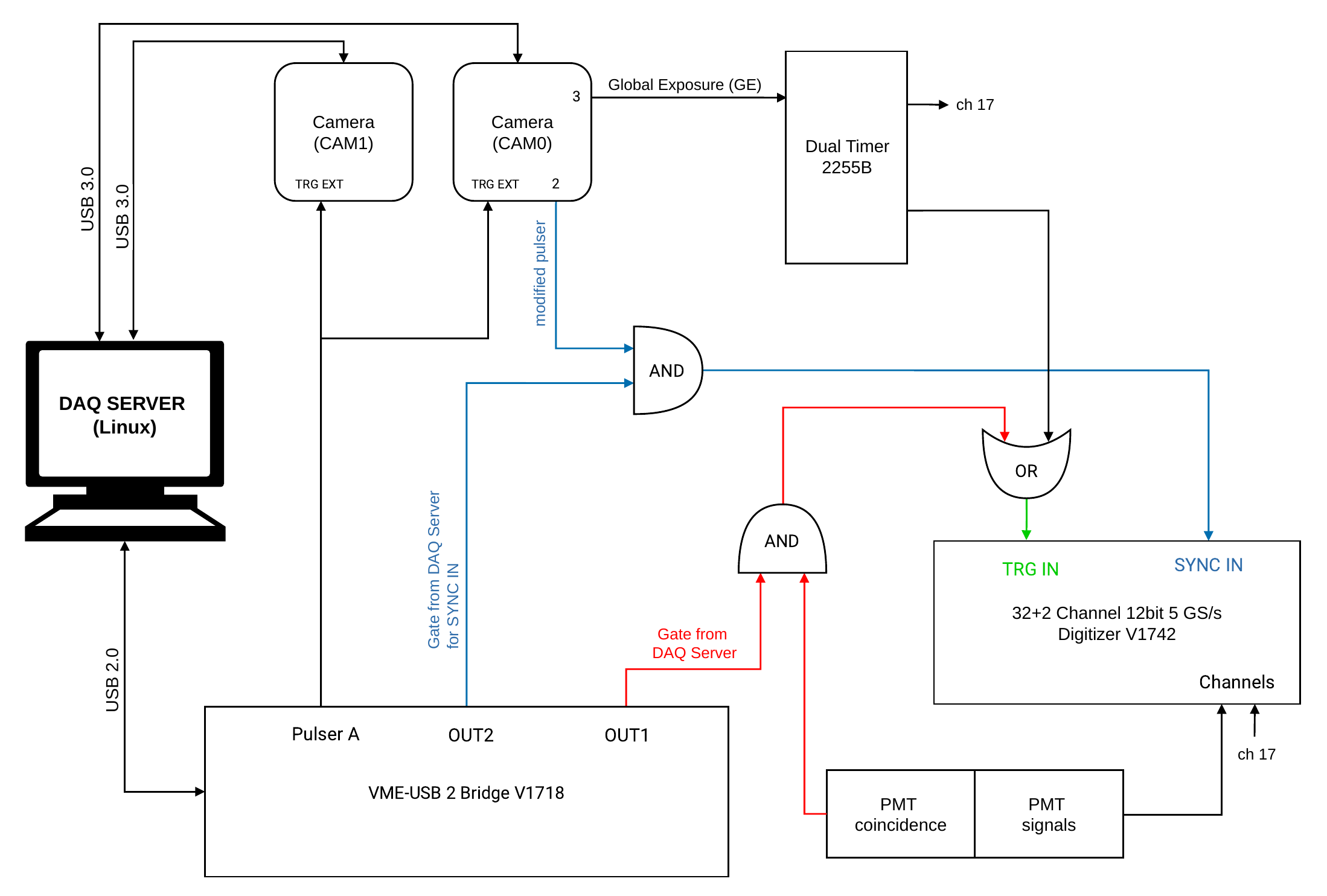}
  \caption{Simplified scheme of the MANGO multi-camera DAQ architecture employing NIM and VME modules.
  The configuration is illustrated with two cameras for clarity, although the design is modular
  and scalable to a larger number of devices.
  The PMT readout chain is omitted, as it follows the same logic adopted in the LIME DAQ.
  The signals contributing to the generation of the \texttt{SYNC-IN} reference for the digitizer
  are highlighted in blue.}
  \label{fig:MangoDAQmultiCAM}
\end{figure}

A schematic overview of the multi-camera DAQ architecture implemented on the MANGO
prototype is shown in Fig.~\ref{fig:MangoDAQmultiCAM}.

For clarity, the scheme is illustrated with two cameras, although the design is modular and can
be readily extended to a larger number of units.
The PMT readout chain is omitted in the figure, as it follows the same logic adopted in the LIME DAQ system.

At the beginning of each run, the DAQ server initiates the timing distribution by configuring a
pulser module, which generates a periodic clock signal.
This signal is split and converted to the \texttt{TTL} logic standard before being distributed to
the external trigger inputs of all cameras.
The cameras operate in synchronous readout trigger mode, whereby the first pulse starts the
exposure of the initial frame and subsequent frames are triggered after a predefined number
of pulses.
The number of pulses is determined by the DAQ server according to the requested exposure
time.

\begin{figure}[h!]
  \centering
  \includegraphics[width=0.95\linewidth]{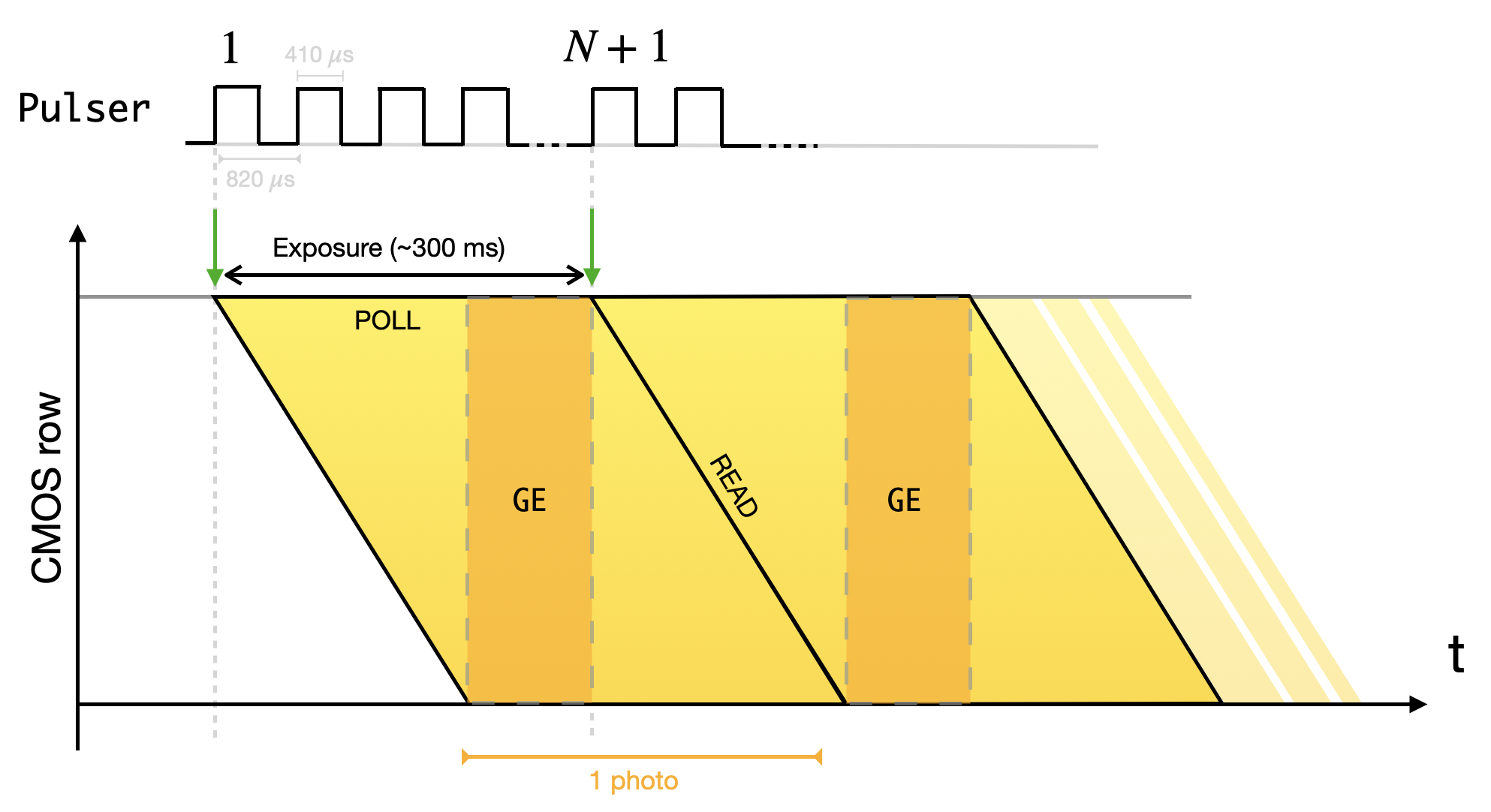}
  \caption{Timing diagram of the cameras operated in synchronous readout trigger mode.
  The first trigger pulse starts the exposure of the initial frame, while subsequent frames are
  triggered after a predefined number of input pulses.
  This scheme ensures that all cameras share a common exposure cadence and aligned frame
  boundaries.}
  \label{fig:NpulserCAM}
\end{figure}

Due to the finite granularity of the pulser, the effective exposure corresponds to the closest
achievable value and is monitored during data taking.
The timing relationship between the input pulses, frame exposure, and readout is illustrated
in Fig.~\ref{fig:NpulserCAM}.
This scheme ensures that all cameras share the same exposure cadence and that frame
boundaries are aligned across devices.

\begin{figure}[]
    \centering
    \includegraphics[width=0.6\linewidth]{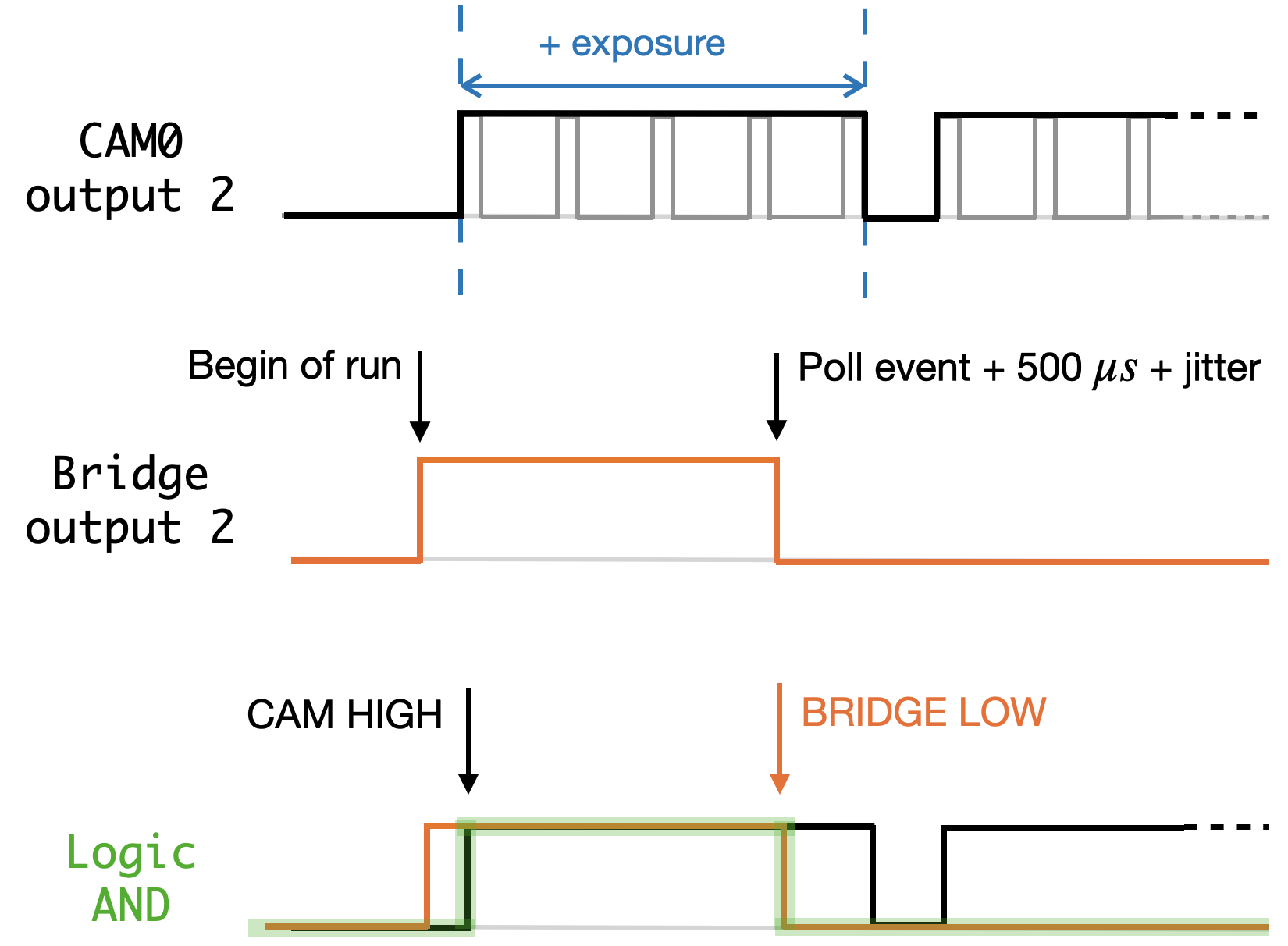}
    \caption{Timing diagram of the signals involved in the generation of the \texttt{SYNC-IN}
reference for the PMT digitizer.
The camera-generated exposure signal defines the rising edge of \texttt{SYNC-IN}, while the
DAQ-controlled gate defines its falling edge, ensuring a well-defined trigger time reference
and preventing false starts.}
    \label{fig: SYNCINmultiCAM}
\end{figure}

The generation of the synchronization signal (\texttt{SYNC-IN}) for the PMT digitizer is derived
from a logical combination of camera and DAQ server signals.
One of the active cameras (\texttt{CAM0} in Fig. \ref{fig:MangoDAQmultiCAM}) provides a modified version of the received pulser signal, in which
the high level is extended for the duration of the exposure.
In parallel, the DAQ server generates, through the VME bridge, a gate signal that is asserted at
the start of the run and released shortly after the DAQ begins polling events.
The logical \texttt{AND} of these two signals directly defines the \texttt{SYNC-IN} reference used
by the digitizer for trigger time-tagging.
In this scheme, the camera-generated signal determines the rising edge of the reference,
while the DAQ-controlled gate defines its falling edge, ensuring a well-defined trigger time tag
origin and preventing false starts.
The timing relationship between the camera-generated signal, the DAQ-controlled gate,
and the resulting \texttt{SYNC-IN} reference is illustrated in Fig.~\ref{fig: SYNCINmultiCAM}.
The PMT trigger logic itself remains unchanged with respect to the LIME configuration.

Image acquisition from each camera is managed independently by the DAQ server, with all
cameras connected via dedicated USB~3.0 interfaces.
The DAQ architecture also includes a dedicated logic to establish a direct timing reference between camera frames and PMT data, which is illustrated in Fig.~\ref{fig:MangoDAQmultiCAM} and described in detail in Sec.~\ref{subsec:camera_pmt_sync}.
The system has been successfully tested with up to three cameras, demonstrating its scalability.
Comprehensive tests with six cameras are foreseen during the pre-commissioning phase of
CYGNO-04.

\subsection{Performance Validation}
\label{subsec:multicamera_validation}

The multi-camera synchronous DAQ architecture was validated with a specific focus on
inter-camera timing alignment and acquisition coherence.
Timing consistency was first assessed by comparing the timestamps associated with frames
acquired by different cameras during synchronous operation.
The distributions of inter-camera timestamp differences, evaluated with a binning of
$1\,\mu\mathrm{s}$, exhibit two nearby peaks separated by approximately $30\,\mu\mathrm{s}$,
which can be attributed to the intrinsic delay introduced during the sensor re-exposure process.

In addition, camera synchronization was directly verified by monitoring the \texttt{Global Exposure}
logical signals from each camera on an oscilloscope.
From the relative positions of the falling edges, an inter-camera timing jitter of the order of
$10\,\mu\mathrm{s}$ was observed.
This value is consistent with the specifications of the Hamamatsu QUEST~2 cameras
\cite{HamamatsuQuest2Manual}.

In addition to the timing alignment, the integrity of the multi-camera acquisition was verified.
No skipped or duplicated frames were observed during synchronous operation at the DAQ level, and the frame indices associated with each camera were found to increase monotonically throughout the runs, indicating correct handling of the acquisition and readout processes.
This behavior was verified on a synchronous acquisition of 500 consecutive frames per camera,
corresponding to a run duration of approximately 3~minutes and 46~seconds.
The synchronization scheme was successfully tested using configurations with two and three
Hamamatsu QUEST~2 cameras, demonstrating the correctness of the approach and its modular
extension beyond a single-camera setup.

Overall, these results confirm that the multi-camera DAQ architecture fulfills the timing and
synchronization requirements foreseen for the CYGNO-04 detector.

\subsection{Synchronization between camera and PMT time coordinates}
\label{subsec:camera_pmt_sync}
A dedicated strategy was implemented to establish a well-defined temporal association between
camera images and PMT signals within the synchronous readout framework.
The goal is to assign an unambiguous trigger time reference to each camera frame that can be
directly compared with the PMT trigger time tags.\\
Within this scheme, the \texttt{Global Exposure} (GE) digital signal from a single camera is used
as the reference for camera timing.
No master camera is defined: any of the synchronized cameras can provide this signal.
The GE signal, whose duration is of the order of the exposure time ($\sim\!100$~ms), is processed by a dual timer module to generate a short logic pulse with a width of approximately
$100$~ns.\\
The resulting pulses, one per acquired frame, are routed to a dedicated channel of the PMT
digitizer and logically combined in \texttt{OR} with the standard PMT coincidence trigger.
This configuration ensures that the camera-related signal is always acquired together with the
PMT waveforms, independently of the presence of a physical trigger.\\
As a result, each camera frame is associated with a distinct and easily identifiable waveform in
the digitizer data, providing a precise trigger time tag for the corresponding image.
Since both the PMT signals and the camera-derived pulses are recorded by the same digitizer,
they share a common time reference.
This guarantees a consistent temporal alignment between optical images and PMT information
at the reconstruction level, without requiring modifications to the PMT trigger logic.

\section{Discussion and Implications for CYGNO-04}
\label{sec:discussion_cygno04}

The T-DAQ upgrades presented in this work were conceived and developed to address key
requirements of the CYGNO-04 detector, most notably continuous imaging and synchronous
multi-camera operation.
These features represent essential building blocks for the next experimental phase of the
CYGNO project, enabling efficient data acquisition from an increased number of optical
readout channels while preserving timing consistency and operational stability.

While the architecture described here provides a validated solution for continuous imaging and
multi-camera synchronization, its direct scaling to the full CYGNO-04 configuration introduces
additional challenges, primarily related to data throughput and storage.
Assuming the standard exposure time of 300~ms adopted in LIME and Hamamatsu QUEST~2 cameras, each camera acquires approximately three images per second.
For a six-camera setup, with an image size of \(2304 \times 4096\) pixels and a depth of
2~bytes per pixel, this corresponds to a raw imaging data rate of approximately
340~MB/s.
Such a sustained write rate is not compatible with long-term data taking and necessitates
the implementation of effective online data-reduction strategies.

The most promising approach to achieve a substantial reduction of the imaging data volume
relies on exploiting the intrinsic sparsity of particle tracks in optical TPC images.
In the CYGNO-04 detector, only a small fraction of pixels in each frame is expected to contain
relevant signal, motivating the development of online event-building and data-selection
techniques capable of identifying and retaining only regions of interest.
By discarding empty or noise-dominated areas at trigger time, these methods can provide
reduction factors well beyond those achievable through coarse event selection alone.

Dedicated algorithms for online region-of-interest identification, including machine-learning-
based approaches, are currently under development within the CYGNO collaboration and are
foreseen to play a central role in the CYGNO-04 DAQ.
Such techniques will be integrated in future iterations of the acquisition system to ensure a
sustainable data rate while preserving the full topological and energy information of physical
events.

Overall, the results presented in this paper demonstrate that the proposed T-DAQ upgrades
provide a robust and flexible foundation for the CYGNO-04 detector.
The validated continuous imaging and multi-camera synchronization schemes address critical
operational requirements and constitute a necessary step toward the realization of a scalable
DAQ architecture for the next phase of the CYGNO experiment.

\section{Conclusions}
\label{sec:conclusions}

In this paper, an upgrade of the CYGNO Trigger and Data Acquisition system has been presented,
addressing key requirements for continuous imaging and synchronous multi-camera operation.
The proposed solutions were designed starting from the LIME T-DAQ and validated on the
MANGO prototype, demonstrating a substantial reduction of dead time, extended trigger time-tagging, and reliable synchronization across multiple cameras.
The developed architecture provides a solid foundation for the CYGNO-04 demonstrator,
enabling efficient operation with an increased number of optical readout channels.
These results constitute an essential step toward the realization of a fully scalable DAQ system
for the next phase of the CYGNO experiment.


\section*{Acknowledgements}

This project has received funding under the European Union’s Horizon 2020 research and innovation program from the European Research Council (ERC) grant agreement No. 818744 and is supported by the Italian Ministry of Education, University and Research through the project PRIN: Progetti di Ricerca di Rilevante Interesse Nazionale “Zero Radioactivity in Future experiment” (Prot. 2017T54J9J). It has also been supported by the PNRR MUR project PE0000013–FAIR. We want to thank General Services and Mechanical Workshops of Laboratori Nazionali di Frascati (LNF). We want to thank the INFN Laboratori Nazionali del Gran Sasso for hosting and supporting the CYGNO project.


\bibliographystyle{unsrt}
\bibliography{bibliography}

\end{document}